\documentclass[prl,twocolumn,showpacs]{revtex4} 
\usepackage{ae}
\usepackage{graphicx} 
\newcommand{\rb}{$^{87}$Rb}

\newcommand{\kq}{$^{41}$K}

\newcommand{\eref}[1]{Eq.~(\ref{#1})}
\newcommand{\fref}[1]{Fig.~\ref{#1}}

\begin{document}

\title{Entropy exchange in a mixture of ultracold atoms}
\author{J.~Catani$^{1,2}$}
\author{G.~Barontini$^{1}$}
\author{G.~Lamporesi$^{1}$}
\author{F.~Rabatti$^{1}$}
\author{G.~Thalhammer$^{1}$}
\author{F.~Minardi$^{1,2}$}
\author{S.~Stringari$^{3}$}
\author{M.~Inguscio$^{1,2}$}
\affiliation{$^1$LENS - European Laboratory for Non-Linear
  Spectroscopy and Dipartimento di Fisica, Universit\`a di Firenze,
  via N. Carrara 1, I-50019 Sesto Fiorentino - Firenze, Italy\\
  $^2$CNR-INFM, via G. Sansone 1, I-50019 Sesto Fiorentino -
  Firenze, Italy\\
  $^3$ Dipartimento di Fisica, Universit\`a di Trento and CNR-INFM BEC
  Center, I-38050 Povo, Trento, Italy}

\begin{abstract}
  We investigate experimentally the entropy transfer between two
  distinguishable atomic quantum gases at ultralow
  temperatures. Exploiting a species-selective trapping potential, we
  are able to control the entropy of one target gas in presence of a
  second auxiliary gas. With this method, we drive the target gas into
  the degenerate regime in conditions of controlled temperature by
  transferring entropy to the auxiliary gas. We envision that our
  method could be useful both to achieve the low entropies required to
  realize new quantum phases and to measure the temperature of atoms
  in deep optical lattices. We verified the thermalization of the two
  species in a 1D lattice.

\end{abstract}


\date{\today}

\maketitle

\begin{figure}[b]
\centering
\includegraphics[width=0.95\columnwidth]{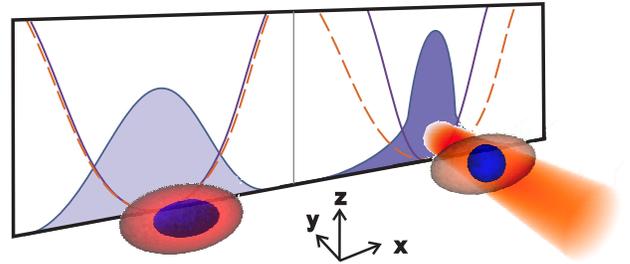}
\caption{(Color online): schematic of our experimental
  procedure. Left: the harmonic magnetic potential is common to both
  gases, auxiliary (red, larger) and target (blue, smaller). Right:
  the species-selective dipole beam compresses the target sample and
  drives it into the degenerate regime. Trapping potentials for the
  auxiliary Rb (dashed) and the target K gas (solid) are sketched on
  the background panels together with the K density distributions.}
\label{fig:sketch}
\end{figure}

In recent years an intense research of quantum phases common to
condensed matter systems and atomic quantum gases has made remarkable
progresses \cite{AdvPhys07-Lewenstein}. Some of these phases can only
be reached provided that the temperature is suitably low. However, in
strongly correlated quantum systems, even the temperature measurement
can be a challenging task. If so, to ascertain whether a given quantum
phase is accessible, it is convenient to focus on the critical value
of entropy, rather than temperature.  The advantage is especially
clear when the strongly correlated regime is reached by sufficiently
slow, entropy-preserving, transformations of the trapping potential,
as it is often the case for atoms in deep optical lattices
\cite{RMP08-Bloch}.  For these reasons, it is important to determine
and grasp control of the entropy of degenerate quantum gases
\cite{PRA06-Cirac, PNAS09-Ho, PRA09-Kohl}. In this work, we
demonstrate the reversible and controlled transfer of entropy between
the two ultracold, harmonically trapped Bose gases, which is based on
the use of a {\it species-selective dipole potential} (SSDP), i.e., an
optical potential experienced exclusively by one species 
%, the target gas (see 
(\fref{fig:sketch}) \cite{PRL03-Presilla, PRA07-Thywissen}. In
particular, we drive the target gas across the threshold for
Bose-Einstein condensation, by a reversible transfer of entropy to the
auxiliary gas.
%, in a reversible manner.

The main idea %is simple and
can be understood from textbook thermodynamics. Let us consider two
distinguishable gases filling an isolated box, exchanging neither
particles nor energy with the outside, and imagine that only one gas
(target) is compressed, e.g. through a piston permeable to the second
gas (auxiliary). The temperature will increase and, in thermal
equilibrium, heat, hence entropy, will transfer from the target to the
auxiliary uncompressed gas. In the limit of the auxiliary gas
containing a large number of particles, it stands as a thermal
bath. In a more formal way, for an ideal gas of $N$ particles, the
entropy $S$ is proportional to $N\log(\Sigma/N)$, where the number of
accessible single-particle states $\Sigma$ increases with the energy
density of states and with the average energy, i.e., the
temperature. In an adiabatic compression of one single gas, the
reduction of the energy density of states is compensated by a
temperature raising such that $\Sigma$, hence $S$, remains
constant. If we add the uncompressed auxiliary gas in thermal contact,
the temperature raising must be lower: $\Sigma$ decreases for the
target gas (and increases for the auxiliary component). In our
experiment the gases are trapped by adjustable harmonic potentials,
but the underlying physics is the same.
% as in the homogeneous case.

To make quantitative predictions we start from the entropy of an ideal
gas at temperature $T$ in a harmonic potential with angular frequency
$\omega$ \cite{BECstringari}: $S=k_B N_{\rm th}
[4g_4(z)/g_3(z)-\log(z)]$, where $N_{\rm th}$ denotes the number of
thermal atoms and the polylogarithmic functions are defined as
$g_n(z)=\sum_{k\geq 1} z^k/k^n$. Above the BEC critical temperature
$T_c$, $N_{\rm th}$ equals the total atom number $N$ and the fugacity
$z$ is implicitily given by the relation
$N=g_3(z)(k_BT/\hbar\omega)^3$.  Below $T_c$, $z=1$ and only the
thermal atoms contribute to the entropy, each with a quantity equal to
%$4k_B\zeta(4)/\zeta(3)\simeq 3.602k_B$, $\zeta(n)=g_n(1)$:
$4k_B\zeta(4)/\zeta(3)$, $\zeta(n)=g_n(1)$, so that
\begin{equation}
  S=4 N k_B \frac{\zeta(4)}{\zeta(3)} \left(\frac{T}{T_c}\right)^3 = 
  4k_B \zeta(4)\left( \frac{k_B T}{\hbar\omega} \right)^3.
  % S/k_B=4 N [\zeta(4) /\zeta(3)](T/T_c)^3 = 4 \zeta(4)(k_B
%   T)^3/(\hbar\omega)^3.
  \label{eq:entropyBEC}
\end{equation}      
Thus, below $T_c$ the entropy is independent of the total number of
atoms $N$. 
% and uniquely determined by the
% temperature and the trapping
% frequencies. 

The above results refer to a %an ideal, i.e.,
non-interacting Bose gas. For high atomic densities, the interatomic
interactions modify low energy spectrum and must be taken into
account: a theoretical analysis is given in
Ref.~\cite{JLTP97-Giorgini}. We evaluate numerically the implicit
expression contained therein and find that, in our experimental
circumstances, the interactions increase the ideal gas values of
entropy by at most 20\% for condensate samples. An expression
for the entropy, easy-to-use and more faithfully approximating the
results of \cite{JLTP97-Giorgini} than \eref{eq:entropyBEC}, is
obtained by replacing $(T/T_c)^3$ in \eref{eq:entropyBEC} with the
thermal fraction $1-f_c$ with $f_c$ taken from a semi-ideal model
\cite{JLTP97-Giorgini, PRA98-Stamperkurn}
\begin{equation}
  \label{eq:fc}
  f_c=1-t^3-\eta(\zeta(2)/\zeta(3))t^2(1-t^3)^{2/5}.
\end{equation}
Here $t=T/T_c$, $\eta=\mu/k_BT$ and the zero-temperature chemical
potential $\mu=(\hbar\omega/2)(15 N a_s/a_{ho})^{2/5}$ depends on
$a_s$, the scattering length, and $a_{ho}=\sqrt{\hbar/m\omega}$. In
our experimental conditions, $\eta$ ranges from 0.30 to 0.33 for
condensed samples.  This approximation is especially helpful for high
condensate fractions ($f_c>0.5$), where the ideal gas formula
\eref{eq:entropyBEC} seriously underestimates the gas
entropy. However, in the explored range $f_c<0.3$,
\eref{eq:entropyBEC} stands as a reasonably good approximation. For
thermal samples above the BEC threshold, the interaction energies are
negligible with respect to the temperature. This is also the case for
interspecies interactions, since the auxiliary sample, always thermal
for the reported data, has low density. For this reason, we simply add
the single-species contributions to obtain the total entropy of the
mixture, $S=S_{\rm K}+S_{\rm Rb}$.

In our experiment, we load a mixture of \rb\ and \kq\ in a millimetric
magnetic trap and sympathetically cool the mixture
\cite{PRA07-DeSarlo}.  The magnetic trap provides a harmonic
confinement with frequencies
$(\omega_x,\omega_y,\omega_z)=2\pi\times(24, 297, 297)$\,Hz for K and
a factor 1.46 smaller for Rb. %Relevant to calculate the entropy is
The geometric mean frequencies
$\omega=(\omega_x\omega_y\omega_z)^{1/3}$ equal to and we have
$(\omega_{\rm Rb},\omega_{\rm K})=2\pi\times(88,128)$\,Hz.  To
selectively act on the potential experienced by the K atoms (target
gas) alone, we use a laser beam tuned to an intermediate wavelength
between the $D_1$ and $D_2$ lines of Rb (auxiliary gas), such that the
dipole forces on Rb due to these two transitions cancel out
\cite{PRA07-Thywissen}. The beam, linearly polarized, with a waist of
$55\,\mu$m, %at the position of the atoms,
propagates along the horizontal $y$ direction (%see
\fref{fig:sketch}). We experimentally determine the %precise
wavelength value to be 789.85(1)\,nm, by minimizing the efficiency of
the Raman-Nath diffraction caused by a pulsed standing wave on a Rb
condensate \cite{PRL99-Phillips-RamanNath}.  We also measured the
residual potential acting on Rb to be $V_{\rm Rb}/V_{\rm K}=
0.08(1)$. Such a residual potential weakly deforms only a small
central region of the Rb density. From the measured temperature
increase caused by the SSDP on Rb alone, we estimate a residual
%compression equivalent to an 
effective increase of $\omega_{\rm Rb}$ lower than
7\%, hereafter neglected.

\begin{figure}[tb]
\centering
\includegraphics[width=\columnwidth]{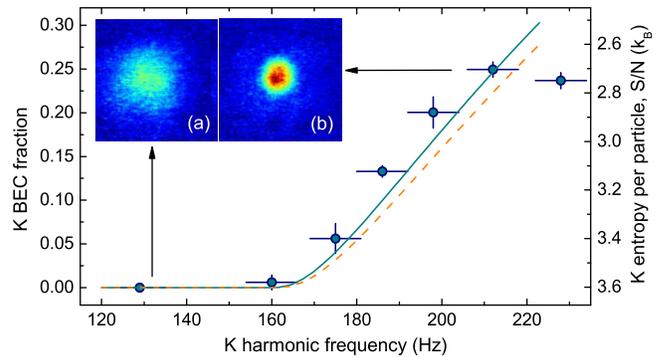}
\caption{(Color online): BEC fraction of the K sample as a function of
  the K harmonic frequency after the compression. Data (circles) are
  compared to the theoretical predictions based on the numerical
  (solid line) and analytical approximate (dashed line) solutions of
  \eref{eq:entropy_eqn}.  On the right axis, the approximated value 
  %single particle entropy valid below BEC threshold, 
$S/(Nk_B)=4(\zeta(4)/\zeta(3))
  (1-f_c)$ is shown. The inset displays absorption images of the K
  sample before (a) and after (b) the compression.}
\label{fig:Kompress}
\end{figure}

On the K sample, the additional confinement induced by the SSDP beam
is instead a harmonic potential whose frequencies add in quadrature to
those of the magnetic trap. The compression occurs mainly along the
weak $x$ axis of the magnetic trap ($\omega_x$ increases up to a
factor 5), slightly along the $z$ direction ($\omega_z$ increases less
than 8\%), and is utterly negligible along the propagation $y$ axis of
the SSDP beam. Overall, the dipole beam raises the $\omega_{\rm K}$, hence
the K critical temperature, up to a factor 1.7.

We raise the power of the SSDP beam with an exponential ramp lasting
200\,ms for maximum compression of the trap frequency $\omega_{\rm
  K,f}=2\pi\times 216$\,Hz. The adiabaticity is preserved since the
ramp is longer than the trapping periods of \kq, i.e.,
(41,3.4,3.4)\,ms. %For lower values of compression, we shorten the
%duration of the ramp at the same time constant. In addition, t
Thermal equilibrium between the two gases is maintained throughout by making
all transformations slow with respect to the interspecies collision
times, a few ms in typical experimental conditions: we verified that
the temperatures of the two species are always equal, within our
statistical uncertainty. Our observables are the number of thermal
atoms $N_{\rm th}$, the number of condensed atoms $N_c$ and the
temperature $T$ of the two species, that are measured by resonant
absorption imaging after all confining potential are removed and the
atomic clouds have expanded. For each species the entropy is either
computed from the measured temperature and atom number for samples
above BEC, or obtained from the condensate fraction below BEC.

In \fref{fig:Kompress} we show that, upon selective compression, the K
sample crosses the BEC threshold, thus its entropy manifestly
drops. Starting with $9.5 \times 10^5$ Rb atoms and $2\times 10^5$ K
atoms at $0.410(15)\,\mu$K, we measure the K condensate fraction at
different compression ratios. 
% A condensate appears when the harmonic
% frequency reaches a critical value of approximately 160\,Hz and
% steadily increases thereafter. 
Due to the large size of the SSDP beam
compared to the K cloud, we observe that, in the absence of Rb, the
compression does not increase the K degeneracy, differently from
%. This behavior is different to that of 
the ``dimple'' configuration, where a tightly
localized dipole trap is used to bring a single species to BEC
\cite{PRL98-Ketterle-Dimple}.

For a comparison of the observed condensate fraction after compression
with the theoretical predictions, we first calculate the final
temperature and then obtain the condensate fraction from
\eref{eq:fc}. The final temperature $T_f$ of the mixture after the
transformation $\omega_{\rm K,i}\rightarrow\,\omega_{\rm K,f}$
satisfies the equation:
\begin{eqnarray}
  \label{eq:entropy_eqn}
  S(T_i, \omega_{\rm K,i}, N_{\rm K})+S(T_i, \omega_{\rm Rb}, N_{\rm Rb}) = & &\\
  = S(T_f, \omega_{\rm K,f}, N_{\rm K})+S(T_f, \omega_{\rm Rb}, N_{\rm Rb}) \nonumber
\end{eqnarray}
where $T_i$ is the initial temperature.  In our experimental
circumstances, the transformation starts with a thermal Rb sample and
a K sample at the BEC threshold or below. Therefore, we solve
\eref{eq:entropy_eqn} numerically for $T_f$. We also find an
analytical approximate solution for $T_f$ \cite{Truncation}:
\begin{equation}
  \label{eq:tfin}
%   T_f=T_i\frac{\omega_{\rm K,f}}{\omega_{\rm K,i}} \left\{\frac{1}{\xi}
%     W\left[\frac{\omega_{\rm K,i}^3}{\omega_{\rm K,f}^3}\xi e^\xi\right]\right\}^{1/3}
  T_f=T_i(\omega_{\rm K,f}/\omega_{\rm K,i}) 
\{W[(\omega_{\rm K,i}/\omega_{\rm K,f})^3\xi e^\xi]/\xi\}^{1/3}
\end{equation}
where $\xi=4\zeta(4)(k_BT_i/\hbar\omega_{\rm K,i})^3/N_{\rm Rb}$ and
$W(y)$ is the Lambert's function defined as the inverse of $y=x e^x$.
In \fref{fig:Kompress} we plot the condensate fraction, and the
related entropy per particle, as a function of the K harmonic
frequency at the end of the compression. The data points are well in
agreement with the numerical results, that differ from the simple
analytical solution obtained from \eref{eq:fc} and (\ref{eq:tfin}) by
only 10\% in our range of compressions.

\begin{figure}[!tb]
\centering
\includegraphics[width=.95\columnwidth]{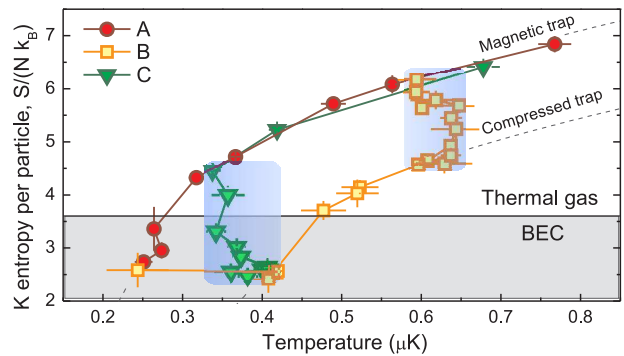}
\caption{(Color online): trajectories in the entropy-temperature
  diagram of the K sample (see text). For the experimental data,
  the entropy is determined with the formulas in text.  The dashed
  lines show the ideal entropy for a sample of $1.1\times 10^5$ K
  atoms at the harmonic frequencies of the magnetic trap 
  ($\omega_{\rm K}=2\pi\times 128$\,Hz) and of the SSDP maximally compressed 
  trap ($\omega_{\rm K}=2\pi \times 216$\,Hz). The shaded boxes highlight
  the isothermal SSDP compressions.}
\label{fig:STdiagram}
\end{figure}

We now illustrate how the interspecies exchange of entropy can be used
to explore the entropy-temperature diagram of a sample of
$1.1(2)\times 10^5$ K atoms, shown in \fref{fig:STdiagram}. Notice
that here we focus on the properties of the target K gas, but the
presence of the auxiliary Rb gas is essential for some
transformations. At each data point, the entropy of the gas is
determined in the following way: above $T_c$, we compute the entropy
per particle by the formula $S/Nk_B=4-\log[N(\hbar\omega/k_B T)^3]$
obtained by truncating the polylogarithmic functions
at the first order in $z$; below $T_c$, we calculate the entropy from
the measured condensate fraction $S/Nk_B=4(\zeta(4)/\zeta(3))(1-f_c)$.

By combining sympathetic cooling and control of the harmonic
frequencies, we can follow different trajectories in the $S$-$T$
diagram. In particular, we compare the efficiency of three
trajectories indicated as A, B, C in \fref{fig:STdiagram}. Trajectory
A corresponds to plain sympathetic cooling in the magnetic trap.
Along B, instead, we raise the SSDP beam to maximum power when the
temperature of the samples reaches $0.6\,\mu$K, and we proceed with
evaporation of Rb and sympathetic cooling of K. When the Rb is
exhausted, at $0.41\,\mu$K we estinguish the SSDP beam intensity, thus
decompressing the trapping potential, and reach the same final point
of the previous trajectory.  Finally along C, we raise the SSDP when
the K sample is close to BEC at $0.34\,\mu$K and cross threshold by
selective compression as in \fref{fig:Kompress}.  The end point of all
trajectories is reached when the Rb sample is either exhausted (B) or
depleted to the point that further sympathetic cooling is inefficient,
$N_{\rm Rb}/N_{\rm K}\sim 2$ (A, C).  The three described paths have
similar efficiencies, since the end values of entropy are
approximately equal. It is important to notice that, for larger
$N_{\rm Rb}/N_{\rm K}$, all cooling processes are more efficient and
lower $S/N$ values are likely attainable.
By means of the SSDP compression in presence of the auxiliary Rb gas, we
perform isothermal transformations. Vice versa, with a single species,
an adiabatic variation of the trapping frequencies corresponds to
an isoentropic transformation that doesn't increase the gas
%any closer to 
degeneracy.  

\begin{figure}[bt]
\centering
\includegraphics[width=.95\columnwidth]{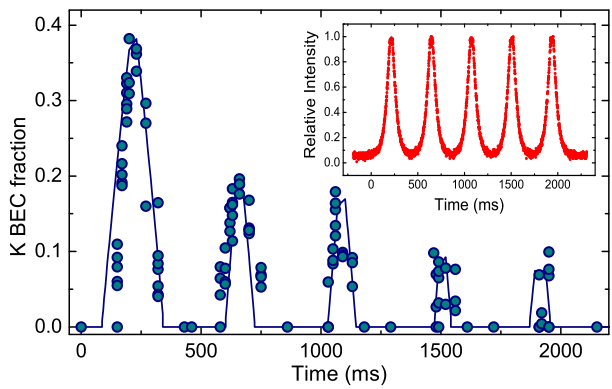}
\caption{(Color online): Cycles of compressions and decompressions,
  the K condensate fraction changes in a reversible manner as we
  modulate the K harmonic frequency over time from 128 to 216\,Hz (the solid line
  guides the eye). The inset shows the relative intensity
  of the SSDP beam.}
\label{fig:cycles}
\end{figure}

The degree of reversibility of the SSDP transformations across the K
BEC threshold is investigated by performing multiple compression and
decompression cycles. The SSDP beam intensity is repeatedly ramped up
and down in an exponential fashion ($\tau=45\,$ms) with a period of
0.43\,s.
As shown in \fref{fig:cycles} the BEC threshold is crossed for
5 cycles at maximum compression. The K condensate fraction
decreases over successive compressions because, at each cycle, the
number of Rb atoms is reduced and the temperature slightly
increased. Even in absence of compression, our Rb sample
experiences an heating rate of the order of $0.7\,\mu$K/s which we
reduce by an order of magnitude by means of a microwave shield
\cite{PRL96-Ketterle}, removing trapped Rb atoms with energy $E/k_B >
5.5\,\mu$K. As a consequence, the Rb atom number decreases at a rate
of $2.5(5)\times 10^5\,$s$^{-1}$. Starting with approximately $8.5
\times 10^5$ Rb atoms and $2 \times 10^5$ K atoms at $0.4\,\mu$K, we
obtain a finite K 
condensate fraction over a time span of 2\,s. After
5 cycles the entropy exchange still occurs but its efficiency is
undermined by the lower Rb atom number and the temperature before the
compression is too high to cross the BEC threshold. Our data can still
be described by the theoretical analysis summarized in \eref{eq:fc}
and (\ref{eq:tfin}), provided we take into account that successive
cycles occur from different initial conditions.

The independent manipulation of two atomic species opens also new
perspectives for the thermometry of cold gases in optical lattices. As
a first application, we have investigated the thermalization of the
two species when the SSDP forms a 1D lattice for K only. The SSDP
standing wave along the $x$ weak axis of the magnetic trap is ramped
in 200\,ms, kept constant for 20\,ms and then linearly extinguished in
1\,ms. The magnetic trap is abruptely switched off immediately
afterwards and the atoms expand freely for 10\,ms (K) and 15\,ms (Rb).
We observe that Rb $x$ and $y$ rms sizes do not depend on the SSDP
lattice depth and either can be used to extract the gas temperature.
We also find that the K temperature obtained from the $y$ size
increases with the lattice strength, but it is always equal to Rb
within our statistical uncertainty of 10\%. This shows that, in a 1D
lattice configuration, up to a strength of $20$ recoil energy, the
auxiliary gas allows to read out the system temperature and to
validate proposed methods to extract the temperature of a gas in an
optical lattice~\cite{PRA09-DeMarco}.

In summary, we have experimentally demonstrated a method to exchange
entropy between two gases, whereby we can precisely reduce the entropy
of an ultracold atomic sample and drive it across the BEC threshold in
a reversible manner.  Earlier experiments demonstrating reversible
BEC, or simply gain in phase-space density, featured either hydrogen
films \cite{PRL97-Walraven}, a single species gas in a ``dimple''
potential \cite{PRL98-Ketterle-Dimple} or multiple spin components in
the same harmonic potential \cite{PRA04-Sengstock}. Our method is
similar in principle to the ``dimple'' configuration of
Ref.~\cite{PRL98-Ketterle-Dimple}, where the single gas can be thought
as consisting of two components distinguished by their extension. The
use of two atomic species combined with the SSDP, however, offers
outstanding flexibility and can be easily extended to different
configurations, e.g. lattices, and to different mixtures.  In
particular, the species-selective potential works best for mixtures
that combine atoms with widely spaced $D_1$ and $D_2$ transitions and
largely different resonant wavelengths.  Our method can be applied
also to reduce the entropy of an ideal degenerate Fermi gas, $S/N
=(\pi^2 k_B^2 T)/(\hbar \omega \sqrt[3]{6 N})$ for harmonic
trapping. In the limit of isothermal transformations, 
%i.e. with a very large auxiliary component, 
the entropy of the target Fermi gas
decreases as the inverse of harmonic frequency compression ratio
$S_f/S_i=\omega_i/\omega_f$, less than in the case of a condensate
$S_f/S_i=(\omega_i/\omega_f)^3$.

A species-selective optical lattice has important applications for the
thermometry of an atomic Mott insulator, filling the need of
convenient experimental techniques \cite{PRA06-Pupillo}. Preliminarly,
we have verified thermalization to occur in a 1D SSDP lattice. In
addition, the auxiliary component could be used as coolant to
dissipate the Mott excitations \cite{PRA04-Daley}. Alternatively,
dilute atoms localized by a species-selective lattice might be used as
disordered scatterers for lattice-insensitive matter
waves~\cite{PRL05-Castin}.
%, as proposed in Ref.

This work was supported by MIUR PRIN 2007, Ente CdR in Firenze, CNR
under project EuroQUAM DQS and EU under STREP CHIMONO and NAME-QUAM.

\bibliography{entropyexchange-r1}

\begin{thebibliography}{21}
\expandafter\ifx\csname natexlab\endcsname\relax\def\natexlab#1{#1}\fi
\expandafter\ifx\csname bibnamefont\endcsname\relax
  \def\bibnamefont#1{#1}\fi
\expandafter\ifx\csname bibfnamefont\endcsname\relax
  \def\bibfnamefont#1{#1}\fi
\expandafter\ifx\csname citenamefont\endcsname\relax
  \def\citenamefont#1{#1}\fi
\expandafter\ifx\csname url\endcsname\relax
  \def\url#1{\texttt{#1}}\fi
\expandafter\ifx\csname urlprefix\endcsname\relax\def\urlprefix{URL }\fi
\providecommand{\bibinfo}[2]{#2}
\providecommand{\eprint}[2][]{\url{#2}}

\bibitem[{\citenamefont{Lewenstein~{\it et al.}}(2007)}]{AdvPhys07-Lewenstein}
\bibinfo{author}{\bibfnamefont{M.}~\bibnamefont{Lewenstein~{\it et al.}}},
  \bibinfo{journal}{Adv. Phys.} \textbf{\bibinfo{volume}{56}},
  \bibinfo{pages}{243} (\bibinfo{year}{2007}).

\bibitem[{\citenamefont{Bloch et~al.}(2008)\citenamefont{Bloch, Dalibard, and
  Zwerger}}]{RMP08-Bloch}
\bibinfo{author}{\bibfnamefont{I.}~\bibnamefont{Bloch}},
  \bibinfo{author}{\bibfnamefont{J.}~\bibnamefont{Dalibard}}, \bibnamefont{and}
  \bibinfo{author}{\bibfnamefont{W.}~\bibnamefont{Zwerger}},
  \bibinfo{journal}{Rev. Mod. Phys.} \textbf{\bibinfo{volume}{80}},
  \bibinfo{pages}{885} (\bibinfo{year}{2008}).

\bibitem[{\citenamefont{Popp~{\it et al.}}(2006)}]{PRA06-Cirac}
\bibinfo{author}{\bibfnamefont{M.}~\bibnamefont{Popp~{\it et al.}}},
  \bibinfo{journal}{Phys. Rev. A} \textbf{\bibinfo{volume}{74}},
  \bibinfo{pages}{013622} (\bibinfo{year}{2006}).

\bibitem[{\citenamefont{Ho and Zhou}(2009)}]{PNAS09-Ho}
\bibinfo{author}{\bibfnamefont{T.-L.} \bibnamefont{Ho}} \bibnamefont{and}
  \bibinfo{author}{\bibfnamefont{Q.}~\bibnamefont{Zhou}},
  \bibinfo{journal}{PNAS} \textbf{\bibinfo{volume}{106}}, \bibinfo{pages}{6916}
  (\bibinfo{year}{2009}).

\bibitem[{\citenamefont{Bernier~{\it et al.}}(2009)}]{PRA09-Kohl}
\bibinfo{author}{\bibfnamefont{J.-S.} \bibnamefont{Bernier~{\it et al.}}},
  \bibinfo{journal}{Phys. Rev. A} \textbf{\bibinfo{volume}{79}},
  \bibinfo{pages}{061601(R)} (\bibinfo{year}{2009}).

\bibitem[{\citenamefont{Presilla and Onofrio}(2003)}]{PRL03-Presilla}
\bibinfo{author}{\bibfnamefont{C.}~\bibnamefont{Presilla}} \bibnamefont{and}
  \bibinfo{author}{\bibfnamefont{R.}~\bibnamefont{Onofrio}},
  \bibinfo{journal}{Phys. Rev. Lett.} \textbf{\bibinfo{volume}{90}},
  \bibinfo{pages}{030404} (\bibinfo{year}{2003}).

\bibitem[{\citenamefont{LeBlanc and Thywissen}(2007)}]{PRA07-Thywissen}
\bibinfo{author}{\bibfnamefont{L.~J.} \bibnamefont{LeBlanc}} \bibnamefont{and}
  \bibinfo{author}{\bibfnamefont{J.~H.} \bibnamefont{Thywissen}},
  \bibinfo{journal}{Phys. Rev. A} \textbf{\bibinfo{volume}{75}},
  \bibinfo{pages}{053612} (\bibinfo{year}{2007}).

\bibitem[{\citenamefont{Pitaevskii and Stringari}(2003)}]{BECstringari}
\bibinfo{author}{\bibfnamefont{L.}~\bibnamefont{Pitaevskii}} \bibnamefont{and}
  \bibinfo{author}{\bibfnamefont{S.}~\bibnamefont{Stringari}},
  \emph{\bibinfo{title}{Bose-Einstein Condensation}}
  (\bibinfo{publisher}{Oxford University Press}, \bibinfo{year}{2003}).

\bibitem[{\citenamefont{Giorgini et~al.}(1997)\citenamefont{Giorgini,
  Pitaevskii, and Stringari}}]{JLTP97-Giorgini}
\bibinfo{author}{\bibfnamefont{S.}~\bibnamefont{Giorgini}},
  \bibinfo{author}{\bibfnamefont{L.~P.} \bibnamefont{Pitaevskii}},
  \bibnamefont{and}
  \bibinfo{author}{\bibfnamefont{S.}~\bibnamefont{Stringari}},
  \bibinfo{journal}{J. Low Temp. Phys.} \textbf{\bibinfo{volume}{109}},
  \bibinfo{pages}{309} (\bibinfo{year}{1997}).

\bibitem[{\citenamefont{Naraschewski and
  Stamper-Kurn}(1998)}]{PRA98-Stamperkurn}
\bibinfo{author}{\bibfnamefont{M.}~\bibnamefont{Naraschewski}}
  \bibnamefont{and} \bibinfo{author}{\bibfnamefont{D.~M.}
  \bibnamefont{Stamper-Kurn}}, \bibinfo{journal}{Phys.\ Rev.\ A}
  \textbf{\bibinfo{volume}{58}}, \bibinfo{pages}{2423} (\bibinfo{year}{1998}).

\bibitem[{\citenamefont{De\ Sarlo~{\it et al.}}(2007)}]{PRA07-DeSarlo}
\bibinfo{author}{\bibfnamefont{L.}~\bibnamefont{De\ Sarlo~{\it et al.}}},
  \bibinfo{journal}{Phys.\ Rev.\ A} \textbf{\bibinfo{volume}{75}},
  \bibinfo{pages}{022715} (\bibinfo{year}{2007}).

\bibitem[{\citenamefont{Ovchinnikov~{\it et
  al.}}(1999)}]{PRL99-Phillips-RamanNath}
\bibinfo{author}{\bibfnamefont{Y.~B.} \bibnamefont{Ovchinnikov~{\it et al.}}},
  \bibinfo{journal}{Phys.\ Rev.\ Lett.} \textbf{\bibinfo{volume}{83}},
  \bibinfo{pages}{284} (\bibinfo{year}{1999}).

\bibitem[{\citenamefont{D.~M. Stamper-Kurn~{\it et
  al.}}(1998)}]{PRL98-Ketterle-Dimple}
\bibinfo{author}{\bibfnamefont{D.~M.} \bibnamefont{D.~M. Stamper-Kurn~{\it et
  al.}}}, \bibinfo{journal}{Phys.\ Rev.\ Lett.} \textbf{\bibinfo{volume}{81}},
  \bibinfo{pages}{2194} (\bibinfo{year}{1998}).

\bibitem[{Tru()}]{Truncation}
\bibinfo{note}{We use $S=4 k_B \zeta(4)(k_B T/\hbar\omega)^3$ for K condensed
  samples and truncate all $z$ series to the linear term so that, for Rb
  thermal samples, $S=N k_B (4-\log[N(\hbar\omega/k_B T)^3]$ .}

\bibitem[{\citenamefont{Mewes~{\it et al.}}(1996)}]{PRL96-Ketterle}
\bibinfo{author}{\bibfnamefont{M.-O.} \bibnamefont{Mewes~{\it et al.}}},
  \bibinfo{journal}{Phys.\ Rev.\ Lett.} \textbf{\bibinfo{volume}{77}},
  \bibinfo{pages}{416} (\bibinfo{year}{1996}).

\bibitem[{\citenamefont{McKay et~al.}(2009)\citenamefont{McKay, White, and
  DeMarco}}]{PRA09-DeMarco}
\bibinfo{author}{\bibfnamefont{D.}~\bibnamefont{McKay}},
  \bibinfo{author}{\bibfnamefont{M.}~\bibnamefont{White}}, \bibnamefont{and}
  \bibinfo{author}{\bibfnamefont{B.}~\bibnamefont{DeMarco}},
  \bibinfo{journal}{Phys. Rev. A} \textbf{\bibinfo{volume}{79}},
  \bibinfo{pages}{063605} (\bibinfo{year}{2009}).

\bibitem[{\citenamefont{Pinkse~{\it et al.}}(1997)}]{PRL97-Walraven}
\bibinfo{author}{\bibfnamefont{P.~W.~H.} \bibnamefont{Pinkse~{\it et al.}}},
  \bibinfo{journal}{Phys. Rev. Lett.} \textbf{\bibinfo{volume}{78}},
  \bibinfo{pages}{990} (\bibinfo{year}{1997}).

\bibitem[{\citenamefont{M.~Erhard~{\it et al.}}(2004)}]{PRA04-Sengstock}
\bibinfo{author}{\bibfnamefont{M.}~\bibnamefont{M.~Erhard~{\it et al.}}},
  \bibinfo{journal}{Phys.\ Rev.\ A} \textbf{\bibinfo{volume}{70}},
  \bibinfo{pages}{031602(R)} (\bibinfo{year}{2004}).

\bibitem[{\citenamefont{Pupillo et~al.}(2006)\citenamefont{Pupillo, Williams,
  and Prokofev}}]{PRA06-Pupillo}
\bibinfo{author}{\bibfnamefont{G.}~\bibnamefont{Pupillo}},
  \bibinfo{author}{\bibfnamefont{C.~J.} \bibnamefont{Williams}},
  \bibnamefont{and} \bibinfo{author}{\bibfnamefont{N.~V.}
  \bibnamefont{Prokofev}}, \bibinfo{journal}{Phys.\ Rev.\ A}
  \textbf{\bibinfo{volume}{73}}, \bibinfo{pages}{013408}
  (\bibinfo{year}{2006}).

\bibitem[{\citenamefont{Daley et~al.}(2004)\citenamefont{Daley, Fedichev, and
  Zoller}}]{PRA04-Daley}
\bibinfo{author}{\bibfnamefont{A.~J.} \bibnamefont{Daley}},
  \bibinfo{author}{\bibfnamefont{P.~O.} \bibnamefont{Fedichev}},
  \bibnamefont{and} \bibinfo{author}{\bibfnamefont{P.}~\bibnamefont{Zoller}},
  \bibinfo{journal}{Phys.\ Rev.\ A} \textbf{\bibinfo{volume}{69}},
  \bibinfo{pages}{022306} (\bibinfo{year}{2004}).

\bibitem[{\citenamefont{Gavish and Castin}(2005)}]{PRL05-Castin}
\bibinfo{author}{\bibfnamefont{U.}~\bibnamefont{Gavish}} \bibnamefont{and}
  \bibinfo{author}{\bibfnamefont{Y.}~\bibnamefont{Castin}},
  \bibinfo{journal}{Phys.\ Rev.\ Lett.} \textbf{\bibinfo{volume}{95}},
  \bibinfo{pages}{020401} (\bibinfo{year}{2005}).

\end{thebibliography}

\end{document}